\documentclass{emulateapj}

\begin{document}

\title{ 
Large Scale Impact of the Cosmological Population of Expanding Radio Galaxies 
}

\author{Paramita Barai\altaffilmark{1}}

\altaffiltext{1}{D\'epartement de physique, de g\'enie physique et d'optique,
Universit\'e Laval, Qu\'ebec, QC, Canada}

\begin{abstract}
We seek to compute the fraction of the volume of the Universe filled by expanding cocoons of 
the cosmological population of radio galaxies over the Hubble time 
as well as the magnetic field infused by them, 
in order to assess their importance in the cosmic evolution of the Universe. 
Using N-body $\Lambda$CDM simulations, 
radio galaxies distributed according to the observed radio luminosity function 
are allowed to evolve in a cosmological volume 
as using well defined prescriptions for their expansion. 
We find that the radio galaxies permeate $10 - 30\%$ of the total volume 
with $\sim 10^{-8}$ G magnetic field by the present epoch. 
\end{abstract}


\section{Introduction} 
\label{sec-intro} 

Radio galaxies (RGs) are believed to have significant impact on the formation and 
evolution of large scale structures in the Universe. 
The cosmological population of expanding RGs and quasars can permeate 
large volumes of the intergalactic medium (IGM) and hence impact a 
considerable fraction of the filamentary protogalactic structures and could 
contribute substantially toward magnetization and metal enrichment of the Universe 
\citep[e.g.,][]{GKW01, kronberg01, FL01, barai04, GWB04, LG05}.

The expansion of shocked and overpressured radio cocoons in 
a two-phase IGM are argued to compress the cold clouds and trigger star 
(perhaps even dwarf galaxy) formation 
\citep{deYoung89, rees89, daly90, chokshi97, natarajan98, vanBreugel04, silk05}, 
as supported by recent observations of jet-induced star formation 
\citep[e.g.,][]{dopita07, reuland07}. 
At the same time, some works indicate that RG expansion inhibits star formation by 
expelling (and heating) the IGM gas 
\citep[e.g.,][]{rawlings04, schawinski06, fujita08}. 
Other studies \citep[e.g.,][]{nath02, vernaleo07, mcNamara07} 
reveal that the RGs heat up the ICM in galaxy clusters. 

A key step to quantify the large scale impact RGs have is 
to address the question that how much of the volume of the Universe do the 
cosmological population of radio cocoons occupy over the Hubble time, 
which we seek to answer in the present work. 

{From} rough calculations, \citet{GKW01} argued that 
the expanding lobes of the generations of RGs can pervade up to 0.5 of the 
WHIM (warm/hot intergalactic medium) component in the 
Universe over $z \sim 1-3$. 
\citet{barai06} performed Monte Carlo simulations to construct 
virtual radio surveys \citep{BRW, wang08}, 
and \citet{barai07} estimated the cumulative volume filling factor to be $\sim0.05$. 
These results were expressed as a fraction of the WHIM volume, 
adopted from the numerical simulations of \citet{cen99}. 

In this work we perform self-consistent cosmological simulations 
to compute the fractional volume of the Universe occupied by RGs, 
a more rigorous approach than previous attempts. 
Finding such volume filling fractions is important 
to probe in more detail the cosmological impact of RGs. 
We also perform preliminary estimates of the magnetic field infused in the filled volumes. 
The simulation method and model are described in \S\ref{sec-model}, 
and the results and discussion are in \S\ref{sec-results}. 

\section{The Model} 
\label{sec-model} 

\subsection{N-body Cosmological Simulation} 
\label{sec-CosmoSim}

We perform N-body simulations of a $\Lambda$CDM Universe, 
where a cubic cosmological box with comoving size $256 h^{-1}$ Mpc on a side, 
having triply periodic boundary conditions and expanding with the Hubble flow, 
is evolved from $z$=25 up to $z$=0. 
The $P^3M$ (particle-particle/particle-mesh) code \citep{hockney81} 
is used with $256^3$ dark matter particles on a $512^3$ grid. 
So a particle has a mass of $1.06 \times 10^{11} M_{\odot}$, 
and the gravitational softening comoving length is $0.3$ of the cell size or $0.15 h^{-1}$ Mpc. 
The cosmological parameters are: present matter density parameter, 
$\Omega_M=0.268$, baryon density parameter, 
$\Omega_b=0.0441$, cosmological constant, 
$\Omega_\Lambda=0.732$, Hubble constant, 
$H_0=70.4\rm\,km\,s^{-1}Mpc^{-1}$ ($h=0.704$), primordial tilt, 
$n_s=0.947$, and CMB temperature, 
$T_{\rm CMB}=2.725$ K, 
consistent with the results of {\sl WMAP3} \citep{spergel07}. 

The baryonic gas distribution is assumed to follow the dark matter in the N-body simulation. 
The ambient gas density, $\rho_x(z, {\bf r})$, is obtained from 
the (matter) density, $\rho_{M}(z, {\bf r})$, 
using $\rho_x = (\Omega_b/\Omega_M) \rho_{M}$. 
The external pressure is then \citep[e.g.,][]{pieri07}, 
$p_x(z, {\bf r}) = \rho_x(z, {\bf r}) k T_x / \mu$. 
The external temperature is fixed at $T_x = 10^4$ K assuming 
a photoheated ambient medium, and $\mu = 0.611$ amu is the mean molecular mass. 
The RGs are distributed in the cosmological volume as given in \S\ref{sec-RLF}. 
They are then allowed to evolve according to the prescription in \S\ref{sec-RGevol}. 
Hence the volume of the cosmological box filled by the expanding RGs is computed. 

\clearpage

\subsection{Initial Source Distribution} 
\label{sec-RLF} 

The cosmological redshift ($z$) and luminosity 
($L \equiv L_{\rm 151~MHz} / {\rm W~Hz}^{-1}~{\rm sr}^{-1}$) distribution of RGs 
is quantified by the radio luminosity function (RLF), which gives the 
number of sources per unit comoving volume per unit $\log_{10}$ of luminosity. 
We adopt the RLF computed by \citet{willott01}, 
modeling it as a combination of two populations: $\rho(L, z) = \rho_l + \rho_h$. 
The low-$L$ population has a number density  
\begin{equation} 
\rho_l = \rho_{l0} \left(\frac{L}{L_{l\star}}\right)^{-\alpha_l} \exp\left(\frac{-L}{L_{l\star}}\right) 
[1 + b(z)]^{k_1}, 
\end{equation} 
where, 
$b(z) = z$ for $z < z_{l0}$, and $b(z) = z_{l0}$ for $z \geq z_{l0}$. 
The number density for the high-$L$ population is 
\begin{equation}
\rho_h = \rho_{h0} \left(\frac{L}{L_{h\star}}\right)^{-\alpha_h} \exp\left(\frac{-L_{h\star}}{L}\right) 
                \exp \left[ -\frac{1}{2} \left( \frac{z-z_{h0}}{\sigma_z} \right)^2 \right], 
\end{equation} 
where, 
$\sigma_z = z_{h1}$ for $z < z_{h0}$, and $\sigma_z = z_{h2}$ for $z \geq z_{h0}$. 
We adopt their model C for the redshift evolution, 
since that is most general with different Gaussian widths at low and high $z$'s. 
The best-fit parameter values for $\Omega_M=0$ and $\Omega_{\Lambda}=0$ are: 
$\log(\rho_{l0})=-7.523$, $\alpha_l=0.586$, $\log(L_{l\star})=26.48$, $z_{l0}=0.710$, 
$k_1=3.48$, $\log(\rho_{h0})=-6.757$, $\alpha_h=2.42$, $\log(L_{h\star})=27.39$, 
$z_{h0}=2.03$, $z_{h1}=0.568$, $z_{h2}=0.956$. 

We convert the RLF to the current consensus cosmology (\S\ref{sec-CosmoSim}) 
using the relation from \citet{peacock85} relating the RLF in two cosmologies, 
$\rho_1(L_1, z) dV_1 / dz = \rho_2(L_2, z) dV_2 / dz$. Then 
\begin{equation}
dN(L, z) = \rho(L, z) ~ d[\log_{10} L] ~ V_{\rm box} 
\end{equation} 
gives the number of RGs within the simulation box of comoving volume 
$V_{\rm box} = (256 h^{-1} {\rm Mpc})^3$ at epoch $z$ in the 
$L$ interval $[L, L+dL]$. 
Sources are generated within radio luminosities 
$24 \leq \log_{10} L \leq 30$. 

The radio luminosity is converted to the constant kinetic power transported by a jet, $Q_0$, 
using the recent result of \citet{koerding07}, 
$\log Q_0 [{\rm erg}/{\rm s}] = 19.1 + \log L_{151} \left[ {\rm W}/({\rm Hz~sr}) \right]$. 

Using the RLF and an assumed active source lifetime, $\tau_{\rm RG}$, we obtain the 
entire cosmological population of RGs in the simulation box starting from $z=8$, 
namely the birth redshift ($z_{\rm bir}$), switch-off redshift ($z_{\rm off}$) and $L$ of each source. 
We implement 3 different values of $\tau_{\rm RG}$: 
10 Myr produces 312340 sources, 100 Myr produces 49282 sources, 
and 500 Myr produces 12807 sources. 

We also consider the possibility that the RG lifetime is inversely proportional to the jet power, 
$\tau_{\rm RG} \propto 1/\sqrt{Q_0}$ \citep{daly02, daly07}, 
where the constant is found assuming a lifetime of 500 Myr at $Q_{0,{\rm min}} = 10^{43.1}$ erg/s. 
This variable $\tau_{\rm RG}$ case produced 47179 sources. 

At each timestep of the simulation, we spatially locate the new RGs 
born during that epoch (whose $z_{\rm bir}$ values fall within the timestep interval) 
in the high-density regions of the cosmological volume. 
We filter the density inside the box such that the RGs are spatially located 
in regions which would collapse to form halos of mass $> 10^{10} M_{\odot}$. 
We consider the mesh cells that have a filtered density $> 5 \times$ the mean density of the box, 
and each new RG is located at the center of one such dense cell, selected randomly. 

\subsection{Radio Galaxy Evolution} 
\label{sec-RGevol} 

After being born, a RG evolves through 
an active-AGN phase (\S\ref{sec-active-AGN}, when $z_{\rm bir} < z < z_{\rm off}$), 
and then goes through a post-AGN phase (\S\ref{sec-dead-AGN}, when $z > z_{\rm off}$). 
At each timestep the total volume occupied by the RGs 
is computed by counting the contributions of all the sources born by then, 
both the active ones and those in the post-AGN phase. 
This gives the redshift evolution of the total RG volume in the simulation box. 

\subsubsection{Active-AGN Phase} 
\label{sec-active-AGN} 

In a RG when the AGN is active, 
relativistic plasma flows down a pair of jets, each of length $R_h$, 
collides with the external environment at the terminal hotspots 
and then flows back to inflate the huge radio cocoon of energetic particles. 
According to the standard scenario \citep[e.g.,][]{begelman89}, 
the cocoon pressure $p_c$ is much larger than the ambient gas pressure. 
The advance speed $v_h$ of the jet head is obtained by balancing 
the jet momentum flux with the ram pressure of the ambient medium, 
\begin{equation} 
\frac{Q_0}{A_h(z) c} = \rho_x(z) v_h^2 (z). 
\end{equation} 
Here $A_h$ is the area of the shocked ``working'' surface at the end of the cocoon 
(larger than the instantaneous cross section area of the jet) and
$\rho_x$ is the external density. 
We use $A_h=2 \pi R_h^2 \theta_h^2$
assuming that the shock front has a constant half-opening angle of 
$\theta_h = 5^{\circ}$ relative to the central AGN \citep{FL01}. 

Since $\tau_{\rm RG}$ is short compared to the Hubble time, 
energy losses and Hubble expansion are neglected in this phase. 
All the kinetic energy transported along the jets during a RG's age 
$t_{\rm age} = t(z) - t(z_{\rm bir})$ is transferred to the cocoon, 
whose energy is $E_c = 2 Q_0 t_{\rm age}$, and 
$p_c V_{\rm RG} = (\Gamma_c-1) E_c$. 
The adiabatic index of the cocoon's plasma, $\Gamma_c =4/3$ as it is taken to be relativistic. 
A shock is driven sideways into the ambient medium at a speed $v_0$, 
following $p_c = \rho_x v_0^2$. 

The RG expands self-similarly during this phase \citep{falle91, KA} 
and we approximate it's shape as cylindrical with length $2R_h$ and radius $R_0$. 
The two equations of motion are: 
$dR_h/dt = v_h$ for advance along the jet, and $dR_0/dt = v_0$ perpendicular to the jet. 
We solve for $R_h$ and $R_0 $ using a 4th-order Runge-Kutta integration method. 
The RG volume during this self-similar expansion  
is then $V_{\rm RG} (z) = 2 \pi R_h R_0 ^2$. 

If $v_0 > v_h$, or, $p_c > \rho_x v_h ^2$, 
the RG loses its self-similarity and ceases to have a cylindrical expansion \citep{begelman89}. 
It then becomes spherical in shape, and the Sedov-Taylor blast wave model 
describing the adiabatic expansion of a hot plasma sphere into a cold medium 
can be used to obtain its radius \citep{castor75, scannapieco04, LG05}, 
\begin{equation} 
\label{eq-Rc} 
R_c (z) = \xi_0 \left( \frac{E_c t_{\rm age}^2} {\overline{\rho_x}(z)} \right) ^ {1/5}.
\end{equation} 
We obtain $\overline{\rho_x} (z)$ by averaging the gas density of the mesh cells 
in the simulation box occurring within the cocoon spherical volume. 
For a strong explosion in the $\Gamma_x = 5/3$ ambient gas, 
$\xi_0 = \{ (75/16\pi) [ (\Gamma_x-1)(\Gamma_x+1)^2 / (3\Gamma_x-1) ] \}^{1/5} = 1.12$ 
\citep{keilty00}. 
Then the RG volume is $V_{\rm RG} (z) = 4 \pi R_c ^3 / 3$. 

\subsubsection{Post-AGN Phase} 
\label{sec-dead-AGN} 

When the AGN activity ends, the cocoon self-similarity is lost 
and we consider that the RG attains a spherical shape, if it had already 
not done so in the active phase. 
The pressure inside the cocoon causes the RG to 
continue expanding as long as it is overpressured \citep{kronberg01, reynolds02}. 
This overpressured cocoon expansion occurs analogous to a 
spherical adiabatic stellar wind bubble, with the radius evolving as in Eq.~(\ref{eq-Rc}). 
Here the total kinetic energy injected into the cocoon by the AGN 
throughout the active RG lifetime is $E_c  =  2 Q_0 \tau_{\rm RG}$. 

The RG is considered to undergo adiabatic expansion losses, and 
the cocoon pressure evolves as $p_c R_c ^{3 \Gamma_x} =$ constant, 
with the constant derived from the pressure and size it had at the end of the active phase. 
The RG follows a spherical expansion as long as its pressure exceeds 
the external pressure, i.e., $p_c(z) > p_x(z)$, and 
in this late expansion phase, $V_{\rm RG} (z) = 4 \pi R_c^3 / 3$. 

When $p_c(z) \leq p_x(z)$, or, the cocoon has reached pressure equilibrium 
with the external medium, the RG has no further intrinsic expansion. 
After this point, the cocoon simply evolves passively with the Hubble flow of the cosmological volume. 
Thus a RG in pressure equilibrium attains a final volume of 
$V_{\rm RG} = 4 \pi R_f^3 / 3$, where $R_f$ is the final comoving radius of the cocoon. 

\section{Results and Discussion} 
\label{sec-results}

\begin{figure} 
\includegraphics[width = 3.2 in]{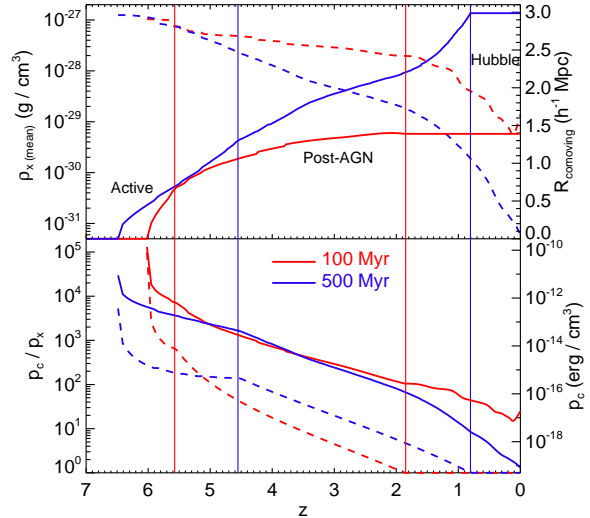}
\caption{
Characteristic quantities for the evolution of two single RGs: 
{\it red} curves for one with $\tau_{\rm RG} = 100$ Myr and $Q_0 = 1.1 \times 10^{44}$ erg/s, 
{\it blue} curves for another with $\tau_{\rm RG} = 500$ Myr and $Q_0 = 1.9 \times 10^{43}$ erg/s. 
Upper panel: 
Comoving size ($R_h$ during active-jet, and $R_c$ during spherical expansion) 
({\it red-solid} and {\it blue-solid} curves with y-axis labels at top-right), 
and 
mean ambient gas density within the cocoon volume ($\overline{\rho_x}$)
({\it red-dashed} and {\it blue-dashed} curves with y-axis labels at top-left). 
Lower panel: 
Cocoon pressure $p_c$ 
({\it red-solid} and {\it blue-solid} curves with y-axis labels at bottom-right), 
and 
overpressure factor of the cocoon w.r.t. external medium $p_c / p_x$ 
({\it red-dashed} and {\it blue-dashed} curves with y-axis labels at bottom-left). 
The vertical lines separate the expansion phases of the RGs: 
active-AGN, post-AGN overpressured and the final passive Hubble evolution.} 
\label{fig1} 
\end{figure}

Figure~\ref{fig1} shows the redshift evolution of two single RGs in the simulation; 
here we discuss only the one with $\tau_{\rm RG} = 100$ Myr ({\it red} curves). 
At the end of the active phase ($z = 5.57$) its cocoon is overpressured by a factor of $\sim650$. 
So it continues to expand while its pressure falls faster because of adiabatic losses. 
Finally when $p_c$ falls to a level to match the external pressure it does not expand anymore. 
From $z = 1.85$ its comoving radius remains constant at $1.4 h^{-1}$ Mpc 
in the passive Hubble phase. 


This illustrates that the radio cocoons are persistently overpressured 
for a substantial period of time even after the AGN has stopped activity, 
and hence continue to expand into the ambient medium. 
In Figure~\ref{fig1}, after a active life of $100$ Myr, the RG remains overpressured for $\sim3000$ Myr. 
Such results are in accord with other studies \citep[e.g.,][]{yamada99, kronberg01}. 
In our simulations $\sim 20 - 50 \%$ (depending on the active lifetime) 
of the sources became spherical in shape during the active-AGN phase. 

In order to prevent overcounting of the volume due to overlap of RGs, 
we count the mesh cells in the simulation box which occur inside the volume 
of one or more RG cocoons. 
The total number of these filled cells, $N_{\rm RG}$, give the total volume of the box occupied by RGs. 
We express the total volume filled as a fraction of volumes of various overdensities in the box, 
$N_{\rho} = N (\rho > {\cal C} \overline{\rho})$, 
where $\overline{\rho} = (1+z)^3 \Omega_M 3H_0^2 / (8 \pi G)$ is the mean matter density 
of a spatially flat Universe (the box) at an epoch $z$. 
So $N_{\rho}$ gives the number of cells which are at a density ${\cal C}$ times the mean density. 
We find $N_{\rho}$ for ${\cal C} = 0, 1, 2, 3, 5, 7$; 
${\cal C} = 0$ gives the total volume of the box, 
since then $N_{\rho} = N (\rho > 0) = 512^3$ is the total number of cells in the box. 

\begin{figure}
\includegraphics[width = 3.2 in]{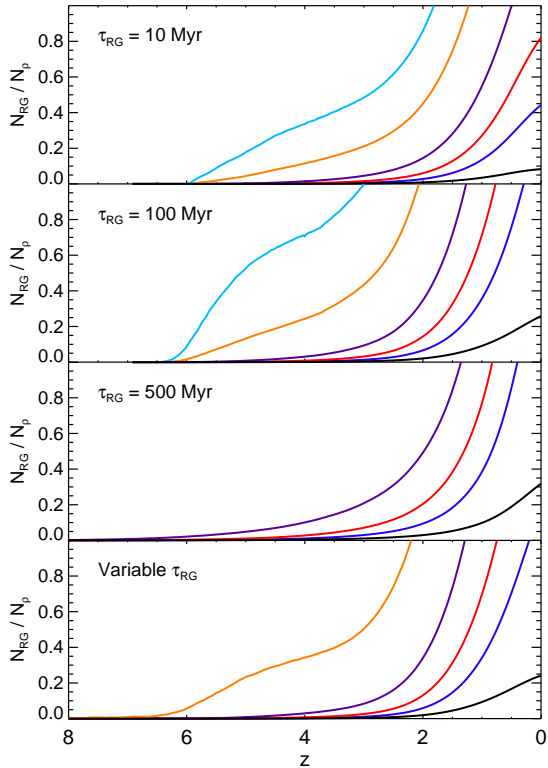}
\caption{Volume filled by RGs ($N_{\rm RG}$) as a fraction of 
total volume of the simulation box {\it (black)}, 
and as a fraction of volumes of various overdensities: 
$N (\rho > \overline{\rho})$ {\it (blue)}, $N (\rho > 2 \overline{\rho})$ {\it (red)}, 
$N (\rho > 3 \overline{\rho})$ {\it (violet)}, $N (\rho > 5 \overline{\rho})$ {\it (orange)}, 
$N (\rho > 7 \overline{\rho})$ {\it (turquoise)}. 
The panels from top to bottom are for active RG lifetimes of $\tau_{\rm RG} = 10, 100, 500$ Myr, 
and for $\tau_{\rm RG} \propto 1/\sqrt{Q_0}$.} 
\label{fig2}
\end{figure}

Figure~\ref{fig2} shows the redshift evolution of the volume filling factors for different active lifetimes. 
By the present epoch, 
$0.08$ of the entire Universe is filled by RGs of active lifetime $10$ Myr, 
the fraction going up to $0.26$ for 100 Myr, and $0.32$ for 500 Myr. 
With $\tau_{\rm RG} = 100$ or $500$ Myr, RGs fill up all of the the overdense regions 
($\rho > \overline{\rho}$, or higher) by $z=0.3-0.4$. 
With $\tau_{\rm RG} = 500$ Myr, 
RGs always fill up regions with $\rho > 5 \overline{\rho}$ or higher at all epochs. 
The case with $\tau_{\rm RG} \propto 1/\sqrt{Q_0}$ fills up $0.24$ of the Universe by $z=0$, 
and give volume filling fractions similar to the values with a constant lifetime of 100 Myr. 


It is the overdense cosmic regions which gravitationally collapse to form stars and galaxies. 
So evidently the RGs have a profound impact on the protogalactic regions of the Universe. 
The precise effect on star formation is still open to debate (\S\ref{sec-intro}), 
with possible RG influence on both triggering and suppressing star formation 
in different regions of the Universe depending on the exact ambient conditions. 

Our volume filling factors of $10-30\%$ are between the values that 
\citet{GKW01} ($50\%$) and \citet{barai07} ($\lesssim 5\%$) obtained as a fraction 
of the volume of the WHIM component of the Universe. 
Our results, based on self-consistent cosmological simulations, 
give a more reliable estimate of the fractional volume of the Universe occupied by RGs. 
The volumes obtained by \citet{LG05} 
(100\% filling by $z\sim1$) are much higher, 
since they consider the whole AGN population. 

We perform preliminary estimates of the energy density and 
magnetic field in the volumes of the Universe filled by radio cocoons. 
The cocoon energy density behaves similar to the cocoon pressure 
evolving adiabatically (\S\ref{sec-dead-AGN}) $u_E = 3 p_c$. 
Assuming equipartition of energy between magnetic field of strength $B_c$ 
and relativistic particles inside the cocoon, 
the magnetic energy density is $u_B = u_E / 2 = B_c^2 / (8 \pi)$. 
The mean thermal energy density of the ambient medium inside the RG volume is 
$\overline{u_{T,x}} = 3 \overline{\rho_x} k T_x / (2 \mu)$. 
We define the volume weighted average of a physical quantity ${\cal A}$ as 
$\langle {\cal A} \rangle (z) \equiv \sum ({\cal A} V_{\rm RG}) / \sum V_{\rm RG}$, 
where the summation is over all RGs existing in the simulation box at that epoch. 


\begin{figure} 
\includegraphics[width = 3.2 in]{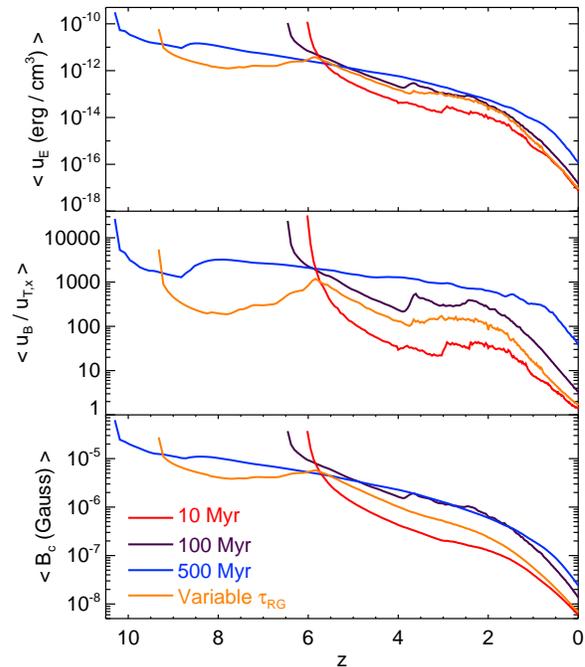} 
\caption{
The volume weighted average of the total energy density 
inside cocoon volumes $\langle u_E \rangle$ (top), 
ratio of the magnetic energy density to the mean external thermal energy density 
$\langle u_B/\overline{u_{T,x}} \rangle$ (middle), 
and the equipartition magnetic field within RG filled volumes $\langle B_c \rangle$ (bottom). 
The color of a curve indicate its lifetime: 
10 Myr ({\it red}), 100 Myr ({\it violet}), 500 Myr ({\it blue}), 
and $\tau_{\rm RG} \propto 1/\sqrt{Q_0}$ ({\it orange}).} 
\label{fig3} 
\end{figure} 

Figure~\ref{fig3} shows the redshift evolution of $\langle u_E \rangle$, 
$\langle u_B/\overline{u_{T,x}} \rangle$ and $\langle B_c \rangle$. 
The energy densities and magnetic field decrease with redshift 
as the filled volumes get bigger. 
The ratio $\langle u_B/\overline{u_{T,x}} \rangle$, giving the importance of cocoon magnetic energy 
over external thermal energy, has a trend similar to that deduced by \citet{FL01}. 
We find that, by the present, $u_B$ is comparable to $\overline{u_{T,x}}$ 
or greater by factors of few, implying that substantial magnetic energies 
are infused into the IGM by the expanding radio cocoons. 
At $z = 0$, a magnetic field of $\sim 10^{-8}$ G permeates the filled volumes, 
consistent with the results of \citet{GKW01} and \citet{ryu98}. 
At a given redshift, 
the energy density and magnetic field are larger for higher source lifetimes. 
The results for $\tau_{\rm RG} \propto 1/\sqrt{Q_0}$ are intermediate 
between those of 10 and 100 Myr. 


We conclude that using our N-body cosmological simulations, 
the expanding population of RGs pervade $10-30\%$ 
of the volume of the Universe by the present, 
and occupy $100\%$ of the overdense regions by $z \sim 0.3$. 
A magnetic field of $\sim 10^{-8}$ G is infused in the filled volumes at $z = 0$.

\acknowledgments 

I thank Hugo Martel for providing me with his $P^3M$ code, and 
him and Paul Wiita for comments on the manuscript and discussions. 
I am grateful to the anonymous referee for helpful comments. 
All calculations were performed at the Laboratoire d'astrophysique num\'erique, Universit\'e Laval. 
I acknowledge support from the Canada Research Chair program and NSERC. 

%




\end{document}